\documentclass[a4papert]{article}
\usepackage[T1]{fontenc}
\usepackage[utf8]{inputenc}
\usepackage{authblk}
\usepackage{fullpage}
\usepackage{amsmath}
\usepackage{amsfonts}
\usepackage{amssymb}
\usepackage{graphicx}
\usepackage[font=small]{caption}
\usepackage{natbib}
\usepackage{tikz}
\usetikzlibrary{trees}
\usepackage{setspace}
\usepackage{subcaption}
\usepackage{longtable}
\usepackage{multirow}
\usepackage{dsfont}
\usepackage{hhline}
\usepackage{bbm}
\usepackage{natbib}

\begin{document}

\author{Ting He}
\affil{School of Finance, Capital University of Economics and Business, Beijing, China}
\title{\Large \bf  Nonparametric Predictive Inference for Asian options}                      
\date{ting.he@cueb.edu.cn}

\maketitle

\centerline{\bf ABSTRACT}
\noindent Asian option, as one of the path-dependent exotic options, is widely traded in the energy market, either for speculation or hedging. However, it is hard to price, especially the one with the arithmetic average price. The traditional trading procedure is either too restrictive by assuming the distribution of the underlying asset or less rigorous by using the approximation. It is attractive to infer the Asian option price with few assumptions of the underlying asset distribution and adopt to the historical data with a nonparametric method. In this paper, we present a novel approach to price the Asian option from an imprecise statistical aspect. Nonparametric Predictive Inference (NPI) is applied to infer the average value of the future underlying asset price, which attempts to make the prediction reflecting more uncertainty because of the limited information. A rational pairwise trading criterion is also proposed in this paper for the Asian options comparison, as a risk measure. The NPI method for the Asian option is illustrated in several examples by using the simulation techniques or the empirical data from the energy market.

\medskip


\clearpage

\section{Introduction}
\doublespacing

Asian options, as one kind of the exotic options, are strongly path-dependent and widely traded in the commodity and foreign exchange market \citep{Kl01}. The main advantages of the Asian option are that its usage of avoiding the risk of market manipulation of the underlying instrument at maturity, and it holds a cheaper price compared to European or American options. The Asian option payoff is contingent on the average value of the underlying asset price, either arithmetic or geometric. For the Asian option settled on the basis of the geometric average price, there are closed formulae by the Black-Scholes model under the assumption that the underlying asset price is the lognormal distributed, so the geometric average price also follows the lognormal distribution with different mean and variance. Although the geometric Asian options are easily priced they are rarely used in practice \citep{Mi98}. While the Asian option with the arithmetic average price is very hard to be evaluated since the density function of the arithmetic average price is unknown\citep{Ve12}. 

Many scholars try to develop and improve the method for the Asian option with the arithmetic average price. One study direction is by assuming  a lognormal diffusion process of the underlying asset price and approximating the density function of the arithmetic average price. The moment matching is used to do the approximation of the option payoff presented by \citet{Tu91}; \citet{Le92}. \citet{Cu94} approximates the payoff of the option by conditioning on the geometric mean price. Another method is to use the numerical method to obtain the solution of the PDE of the Asian option.  The problem of this method is that when the explicit finite difference method is used in PDE of a path-dependent option pricing, it is numerically unstable. The implicit method is stable referring to the Asian option pricing, but it only provides the result for some specific volatility structure. \citet{Ve01} improves the instability problem by presenting a numerical one dimensional PDE for the Asian option pricing which is stable under the finite difference method. Monte Carlo simulation \citep{Bo77} as a very effective way to price the path-dependent option that has been developed for the Asian option pricing \citep{Hu09}. \citet{Ke90} present a Monte Carlo strategy of pricing the option with the arithmetic average price with the variance reduction elements. \citet{Ba13} study the Asian option pricing problem by presenting a joint Monte Carlo-Fourier transform sampling scheme under the CGMY process. The concern of Monte Carlo simulation of the option pricing is to estimate an accurate result is very time-consuming. Another popular method is the discrete lattice method. \citet{Hu93} propose the first tree pricing model for Asian options, which has some drawbacks of the approximation precision and the convergence to the continuous value. \citet{Kl01} and \citet{Da08} improve Hull and White's tree model considering the representative average prices to limit the approximation error.  \citet{Li14} present a binomial approach for the Asian option pricing leading to the upper and lower bounds of the approximation result reducing the interpolation error. The study discussed above is based on the assumption that the future pattern of the Asian option is well known. When the information of the future market is limited, imprecise probability allows us to predict the Asian option price with the observed information.

Imprecise probability as a generalization of classical probability theory enables various less restrictive representations of uncertainty \citep{Au14}. Nonparametric predictive inference (NPI) is one of the statistical inference methods for imprecise probability, which is a frequentist statistics framework with strong consistency properties \citep{Au04}; \citep{Co11}.  The NPI method provides the approach to calculate the upper and lower probabilities of the interested event aiming to do the prediction by making few assumptions in addition to observed data. One property of the NPI method is when multiple future observations are predicted, the observations are interdependent, meaning after one prediction, this predicted value is added to the observed data together forecasting the next future observations. Therefore, the NPI method reflects more uncertainty by increasing the variability if the multiple future observations are assumed to be conditionally independent. The NPI method has been applied to the finance area, predicting future stock returns when little further information is available and providing a way of the pairwise comparison between stock returns \citep{Ba17}. \citet{Co18} presented a new approach to quantify the credit risk by using the NPI method based on the ROC analysis. The implements of the NPI method for the vanilla option pricing well perform when there are less certain information of the underlying asset \citep{He19};\citep{He20}.

In this paper, we present the NPI method for Asian option pricing, which attempts to evaluate the Asian option price based on the historical data and offering a rational pairwise trading scenario. Some relevant background about the NPI method is summarized in Section 2. The Asian option pricing model based on the NPI method is proposed in Section 3 along with a rational trading criterion of the comparison of two Asian options. In Section 4, we illustrate the NPI method by using the simulation as well as the empirical examples of the energy market. Some conclusions and extensions are written in Section 5.

\section{Preliminaries}

Nonparametric Predictive Inference (NPI) is an inferential framework based on the assumption $A_{(n)}$  \citep{Hi68}, which directly provides probabilities for future observations by using few model assumptions and observed values of related random quantities.  $A_{(n)}$ assumption makes sure that the future observation is equally likely to fall in the interval of a real value line created by $n$ observed random quantities, which keeps the consistency of the exchangeability. Based on the $A_{(n)}$ assumption, NPI offers the upper and lower probabilities for one or multiple future random quantities when $n$ observed random quantities are available, which follows De Finetti's fundamental theorem of probability \citep{De74}. NPI is a frequentist statistical method which has strong consistency properties \citep{Au04}. 

NPI for $m$ multiple future random quantities is concerned in this paper, which is based on  $A_{(n+m-1)}$. The NPI method predicts the future observation based on historical data and keeps updating the data, which means the prediction of $m$ future data is identical to the prediction of one future data. After the first one is predicted, the prediction is used as the historical information to forecast the next observation. NPI assumes that each future data is equally likely in the interval $I_j$, where $j=1,2,...,n+1$, created by $n$ observed data, which also means that all possible orderings of $n$ observed data and $m$ future data are equally likely. Totally, ${n+m \choose m}$ possible orderings can be derived, so for each ordering the probability is equal to ${n+m \choose m}^{-1}$.

To inference the arithmetic average price of the underlying asset based on the NPI method, we first forcast the return of the underlying asset.  Baker  \citep{Ba17} predicts the stock future return with few information by applying the NPI method, but the predicted return is under the frame of simple interest. In this paper, we follows the same idea and extend it to compound interest. Through the prediction of the underlying asset return, the imprecise arithmeric price of the underlying asset can inferred and utilized in the Asian option pricing procedure.

\section{NPI for Asian options}
In this section, an Asian option pricing method is presented based on the NPI method, which  reflects the uncertainty not only from the stochastic environment but also from the limited prior information. And a trading criterion by comparing the Asian options contingent on two different product is shown as a risk measure. 
\subsection{Prediction stock returns }
Define $S_t$ is the underlying asset price at time $t$. By assuming there are $n$ historical underlying asset price $S_t=s_t, t=1,2,\ldots,n$ available in the market, where the time intervals between these historical data are identical to each other. Then the continuous compounding rates of return of the underlying asset price $r_t$ is,
$$r_t=ln\left(\frac{S_t}{S_{t+1}}\right), \ t=1,2,\ldots,n$$
To predict future return of underlying asset price based on the NPI method,  the exchangeability is assumed in our model meaning the order of the underlying asset return is irrelevant. After we calculate the compounding return, we rank these values from the lowest value to the highest value, $r_{(1)}, \ldots, r$  $_{(n)}$. Then on this real value line created by $r_{(1)}, \ldots, r_{(n)}$, there are $n+1$ intervals. To avoid the influences of $\infty$ and $-\infty$, we need to find the lowest and the highest returns, $r_{(0)}$ and $r_{(n+1)}$, which can be the extreme returns in a long-term historical period or the extreme values referring to the user's preference. On the basis of the historical information, we assume that the future data randomly falls in any interval on this real value line. From the assumption of   multiple future data prediction through the NPI method, totally there are ${n+m \choose m}$ orderings of $m$ future compounding returns, which are equally likely. Investor can infer the future returns by counting the orderings fitted in one's investment criterion.

As the aim of this paper is to study the Asian option with arithmetic average and do the prediction, the aggregate compounding return is concerned. The general formula of the aggregate compounding return is
$$\hat{R_i}=\frac{\sum_{t=n+1}^{n+i}R_t}{i}$$
This presents the aggregate compounding return for $i$ future cumulative time, where $i=1,\ldots,m$. For example, when $i$ is equal to one, $\hat{R_1}$ represents the aggregate return during the first time step, and when $i=2$, $\hat{R_1}$ represents the aggregate return during the first and second time steps and so on. By applying this formula to the NPI framework, the upper bound $\hat{R_i^u}$ and the lower bound  $\hat{R_i^l}$ of the aggregate compounding return for $i$ future period can be calculated. The fundamental idea is that for a specific ordering, $R_t$ randomly falls in the interval $I_j$, $j=1,\ldots,n+1$, which defines the upper bound $R_t^u$ of $R_t$ equal to $r_{(j)}$ and the lower bound  $R_t^l$ equal to $r_{(j-1)}$. By putting the upper and lower bounds of $R_t$ into the aggregate calculation, the upper and lower bounds of $\hat{R_i}$ can be obtained.

\subsection{ Asian option expected prices based on the NPI method}
As we mentioned, the Asian option's payoff depends on two type of average value, the arithmetic average price and the geometric average price. In this paper, the Asian option with arithmetic average price is in consideration. According to the different type of the strike price , there are two types of Asian options, with the fixed strike price and the float price. The Asian option with fixed strike price is discussed in this paper. Therefore, if a $m$ period Asian option with fixed strike price $K$ is priced, the general pricing formula is,
\begin{equation}
\label{eq:AsianF}
V_0=B(0,m)[S_\mu^m-K]^+
\end{equation}
where $V_0$ is the initial expected price of this Asian option, $S_\mu^m$ is the arithmetic average price of the underlying asset during $m$ period, and $B(0,m)$ is the discount factor during $m$ period. From Equation (\ref{eq:AsianF}), we can conclude that the payoff of this type Asian option is the positive value of the subtraction between the average underlying asset price and the predetermined strike price.

To calculate the arithmetic average price of the underlying asset during $m$ period, the aggregate compounding returns for every $i\in(1,\ldots,n)$ period are needed. 
\begin{equation}
\label{eq:AthS}
S_\mu^m=\frac{1}{m}\sum_{k=0}^mS_0e^{i\hat{R_i}}
\end{equation}
where $S_0$ is the initial underlying asset price and $R_0$ is set to be zero. By the definition of the Asian option, the exact value of the future underlying asset is less important. Rather than the explicit value of each time step $S_t$, the average behavior of the underlying asset is considered, where the aggregate return is the appropriate value to represent the asset behavior during a period. Thus, based on the NPI method, we do not concern about the exact value of $S_{1}, \ldots, S_{m}$. Instead, the upper and lower bounds of aggregate compounding returns for every $i\in(1,\ldots,n)$ time-steps are calculated. Putting the bounds of the compounding returns in Equation (\ref{eq:AthS}), we get the upper and lower bounds of the arithmetic average underlying asset price. For $m$ future time steps, the minimum average underlying asset price is,

\begin{equation}\label{eq:2.6}
\underline{S_\mu^m}=\frac{1}{m}\sum_{k=0}^mS_0e^{i\hat{R_i^l}}
\end{equation}

The maximum average underlying asset is,
\begin{equation}\label{eq:2.7}
\overline{S_\mu^m}=\frac{1}{m}\sum_{k=0}^mS_0e^{i\hat{R_i^u}}
\end{equation}

Thus, according to the definition of the Asian option payoff, we can calculate the upper and lower expected option price based on the NPI method, which is called the minimum selling price and the maximum buying price according to the trading intention.

\textbf{The minimum selling price for the call option}
\begin{equation}\label{eq:NPI1}
\overline{V_0}=B(0,m)[\overline{S_\mu^m}-K]^+=B(0,m)\left[ \frac{1}{m}\sum_{k=0}^mS_0e^{i\hat{R_i^l}}-K\right] ^+
\end{equation}

\textbf{The maximum buying price for the call option}
\begin{equation}\label{eq:NPI2}
\underline{V_0}=B(0,m)[\underline{S_\mu^m}-K]^+=B(0,m)\left[ \frac{1}{m}\sum_{k=0}^mS_0e^{i\hat{R_i^u}}-K\right] ^+
\end{equation}

\textbf{The minimum selling price for the put option}
\begin{equation}\label{eq:NPI3}
\overline{V_0}=B(0,m)[K-\overline{S_\mu^m}]^+=B(0,m)\left[ K-\frac{1}{m}\sum_{k=0}^mS_0e^{i\hat{R_i^u}}\right] ^+
\end{equation}

\textbf{The maximum buying price for the put option}
\begin{equation}\label{eq:NPI4}
\underline{V_0}=B(0,m)[K-\underline{S_\mu^m}]^+=B(0,m)\left[ K-\frac{1}{m}\sum_{k=0}^mS_0e^{i\hat{R_i^l}}\right] ^+
\end{equation}

The upper and lower bounds of the Asian option indicate the buying and selling thresholds  of the investor. The investor who trades according to the result from the NPI method would not like to be in the game when the quoted price is in the interval of the minimum selling price and the maximum buying price. However, if the quoted price is higher than the minimum selling price, the investor prefers to sell the option. Or the longing position is triggered when the maximum buying price is greater than the quoted price. The advantage of this method is we do not assume any distribution of the underlying asset distribution. The prediction is based on the information from the historical data. Different from calculating the average price of historical data directly, this method considers the randomness of the stock price and its outcome is an interval, which avoids the error of the historical data bias and reflects more uncertainty of the underlying asset.

\subsection{Trading criteria of the Asian options contingent on two underlying asset}\label{s:2.2}
Other than pricing the Asian option by using the NPI method, this method also offers a way to make a reasonable decision in the Asian option trade. The NPI method provides the upper and lower probability that the investor can get a positive profit in this investment. Suppose there is a sequence of historical data with an amount of $n$, which is continuous, consistent and exchangeable. Same as what we did for the option pricing procedure, we calculate the historical aggregate compounding returns and rank them from the lowest one to the highest one  $r_{(1)},r_{(2)}...,r_{(n)}$. And to avoid the influence from the infinity values, we set the new historical sequence started with $r_{(0)}$, and end with $ r_{(n+1)}$, which these two values can be determined by using the minimum historical price and the maximum historical price in a long-term time. By having the aggregate compounding returns, we can calculate the average price of the underlying asset $S_\mu^m$. Next, the NPI lower and upper probabilities of the positive payoff are derived for the Asian option  involving the average stock price $S_\mu^m$ and the strike Price $K$. The investor can use these probabilities to compare the Asian options and set their trading criteria. The formulae are listed below.

\textbf{The upper and lower probability of a positive payoff} 
\begin{equation}
\overline{P}(\text{Payoff}>0)=
\begin{cases}
\frac{1}{{n+m\choose m}}\sum_O \mathbbm{1} \lbrace \overline{S_\mu^m}>K\rbrace & \text{call option}\\
\frac{1}{{n+m\choose m}}\sum_O \mathbbm{1} \lbrace \underline{S_\mu^m}<K\rbrace & \text{put option}
\end{cases}
\end{equation}
\begin{equation}
\underline{P}(\text{Payoff}>0)=
\begin{cases}
\frac{1}{{n+m\choose m}}\sum_O\mathbbm{1}\lbrace \underline{S_\mu^m}>K\rbrace & \text{call option}\\
\frac{1}{{n+m\choose m}}\sum_O \mathbbm{1} \lbrace \overline{S_\mu^m}<K\rbrace & \text{put option}
\end{cases}
\end{equation}

where$\overline{S_\mu^m}$, $\underline{S_\mu^m}$ are calculated by Equations (\ref{eq:2.6}) and (\ref{eq:2.7}). $\sum_O$ is the summation over all the$\frac{1}{{n+m\choose m}}$possible orderings of the m future returns within the $n+1$ intervals, and $\mathbbm{1}\lbrace A \rbrace $ is an indicator function which is equal to 1 if $A$ is true or 0 otherwise.

This interval probabilities can help an investor to choose the better underlying asset in the Asian option investment as either a speculator or a hedger. As in the commodity market,  especially in the crude oil market. there are a variety of underlying assets correlated to each other, so it is hard to choose which underlying asset is a better investment. By the NPI method, an investor is offered an indicator that can be referred according to the investor's risk aversion and character, speculator or hedger.  Suppose there are two underlying assets $A$ and $B$ that have similar price values and trends. A speculator is a risk-taker whose purpose of an investment is to seek the opportunities to earn some profit. If the speculator would like to invest in either of these two assets, then the indicator below suggests the speculator invest in asset $A$, the lower probability of a positive payoff for asset $A$, $\underline{P}(\text{Payoff}_A)$, is greater than the lower probability of a positive payoff for asset $B$, $\underline{P}(\text{Payoff}_B)$, or the upper probability of a positive payoff for asset $A$, $\overline{P}(\text{Payoff}_A)$, is greater than the upper probability of a positive payoff for asset $B$, $\overline{P}(\text{Payoff}_B)$. For a hedger, the purpose involving in the option trading is to hedge the risk in the trade of the underlying asset, so the hedger has a high level of risk aversion. An absolute strength of asset $A$ needs to be revealed to instruct this hedger's action.  Thus, when $\underline{P}(\text{Payoff}_A)>\overline{P}(\text{Payoff}_B)$, the investment in asset $A$ is appealing to the hedger.

\section{Illustrated examples}
Several examples are discussed in this section to illustrate the NPI method for the Asian option. We first study the performance of the NPI method for Asian option pricing by the simulation techniques. Then a performance study of the energy market is developed to assess the empirical value of this method.
\subsection{The simulation study}

\begin{figure}
\centering
\includegraphics[scale=0.8]{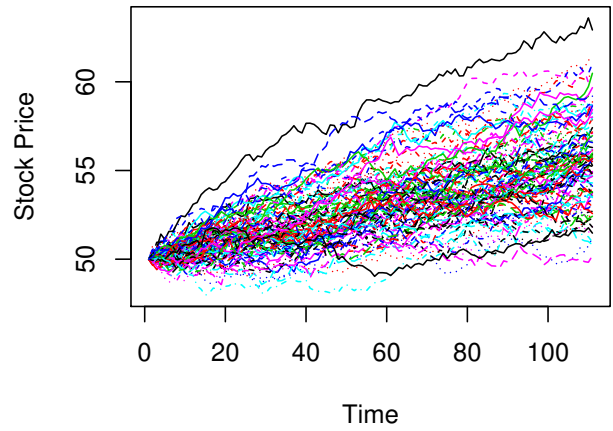}
\caption{Simulated stock price paths}\label{fg:GBM}
\end{figure}

As acknowledged, the Geometric Brownian Motion (GBM)  is widely used to model the stock price behavior. Therefore, to start the illustration, an example based on the GBM is presented in this section. By utilizing the R program, we first simulate 100 paths of stock prices following the GBM with the return equal to 0.02 and the volatility equal to 0.02 as well. The simulated stock paths are displayed in Figure \ref{fg:GBM}. In each path, the initial price is 50, and the program simulated the stock price movement for 110 time steps. In our example, the time period 0 to 100, is assumed as the historical time period calling  the corresponding data the historical data, while assuming the time period 100 to 110, to be the predictive time period calling the corresponding data the future data. The idea is using the NPI method for the Asian call option pricing formulae, Equations (\ref{eq:NPI1}) and (\ref{eq:NPI2}) to forecast the option price, and using the future data to calculate the option price based on payoff definition as the benchmark. In the following example, we predict the price of an at-the-money (ATM) call option where the strike price equals to the initial price  50.

\begin{figure}
\centering
\includegraphics[scale=0.8]{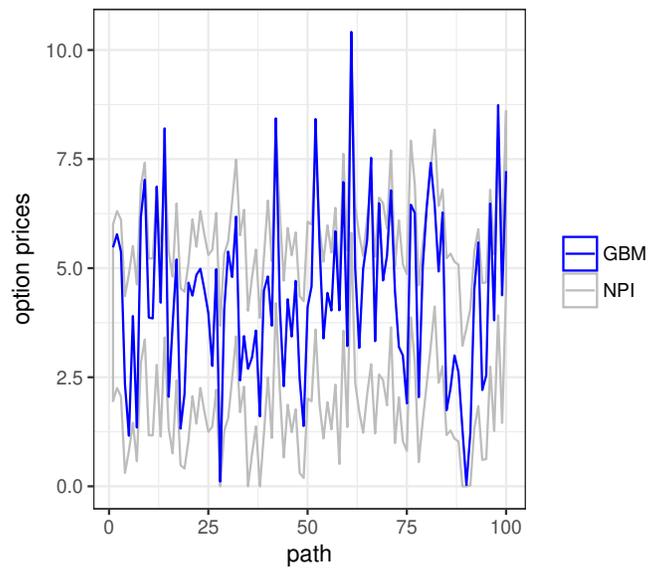}
\caption{Asian option prices predicted by the NPI method with the  GBM option price as the benchmark}\label{fg:AS_NPI_GBM}
\end{figure}

Figure \ref{fg:AS_NPI_GBM} discloses that the NPI method can provide an interval that includes the benchmark price in most of the cases. In these cases, the NPI method's prediction includes the benchmark value, but this does not mean that NPI method can predict the price accurately. If the result from the NPI method is an interval with a large value gap, the benchmark price could be in the interval for sure. Some further discussions  of the accuracy based on the NPI method are illustrated in the next paragraph.  Figure \ref{fg:AS_NPI_GBM} also reveals that the fluctuations of NPI option prices have a similar pattern to that of the GMB option prices. In these 100 paths, for the GBM option price with a higher value, the maximum buying and minimum selling prices corresponding to this path also have a higher value, and the same conclude can be derived from the figure for the path with a lower GBM option price as well. However, it is also clear that the boundary prices from the NPI method are less fluctuated than the GBM prices. It is easy to understand from the perspective of deviation. As the GBM prices are predicted based on future data, it has a greater deviation from the initial price of the simulation than that of the historical data, while the NPI method derives the option prices from the historical data. So the patterns of NPI option boundary prices are more stable than the pattern of the GBM prices. Therefore, there are some extreme values manifested in the GBM price. But the corresponding extreme values in the NPI option boundary prices are less significant. 

To investigate the influence of the volatility on the prediction outcome from the NPI method, we define three factors; coverage percentage, accuracy and precision. Coverage percentage estimates the percentage of NPI outcomes including the benchmark value in the interval. Accuracy is defined as the expectation of absolute difference between the median value of the NPI interval and the benchmark value, $E[|\text{median}_{\text{NPI}}-\text{benchmark}_{\text{GBM}}|]$, which reflects the deviation between the NPI outcomes and the benchmark. Precision is to calculate the mean value of the interval length from the NPI method. Including the precision in our study is because if the precision is very large, the result of the coverage percentage is supposed to be better than the case when the precision is very small. Herein, we study the influence of the varying volatilities on these three factors in order to estimate the performance of the NPI prediction result for the same ATM option as that in the last example. In this study, the volatility is in the range from 5\% to 10\% to simulate the daily volatility in the market. Three factors are monitored to assess the NPI results. 

\begin{longtable}[tp]{ |p{1.8cm}||p{1.8cm}|p{1.8cm}||p{1.8cm}|p{1.8cm}|}
\hline
&\multicolumn{2}{c||}{Precision is large [3,4]}&\multicolumn{2}{c|}{Precision is small[1,1.5]}\\
\hline
volatility&percentage&accuracy&percentage&accuracy\\
\hline
0.5\%&1&0.5856104&0.9653&1.391795\\
\hline
1\%&0.9999&0.6747436&0.9625&1.401927\\
\hline
1.5\%&0.9963&0.8498949&0.9576&1.421381\\
\hline
2\% &0.9828 & 1.047153   & 0.9597  &1.501727 \\
\hline
2.5\%&0.951  &1.262822 &0.9345 &1.609755\\
\hline
3\% &0.9208 & 1.425547 &0.8975 & 1.72399\\
\hline
3.5\% &0.8797 &1.590992  &0.8744 &1.866814\\
\hline
4\% &0.8481 &1.752489  &0.8289&1.969898\\
\hline
4.5\%&0.8271 &1.937935  & 0.8001 & 2.082382 \\
\hline
5\% &0.7876 &2.064087  &0.7875&2.188028 \\
\hline
5.5\%&0.7638 &2.174651 & 0.7576&2.358233 \\
\hline
6\% &0.7387 & 2.301093  & 0.7376 &2.47124\\
\hline
6.5\%&0.7169 &2.459455  &0.7137 & 2.554012 \\
\hline
7\% &0.6921 &2.543524 &0.6897&2.7054 \\
\hline
7.5\% &0.6832 &2.706887  & 0.6722&2.819131 \\
\hline
8\% &0.6808 &2.821586 & 0.6625 &2.980532 \\
\hline
8.5\% &0.6514 &2.913285  &0.6504 &3.059432 \\
\hline
9\% &0.6403 &3.012923  &0.64 &3.16712 \\
\hline
9.5\% &0.6364 &3.167188  &0.6242 &3.296574\\
\hline
10\%&0.6153&3.283862&0.6147&3.356551\\
\hline
\caption{The study of volatility influence}\label{ta:3factors}
  \end{longtable}
 
Table \ref{ta:3factors} displays the outcomes of three factors with the varying volatility in two simulations. When we calculate the precision with different volatilities, we find that as the volatility increases, the precision decreases. This result does not mean that high volatility has a positive effect on the precision. The reason why high volatility causes a small precision is when the underlying asset is more volatile,  there are more times of the simulated paths with a zero payoff either from NPI method or from the GBM model. As the average precision value of all simulated paths is calculated as the estimator, the average result is getting smaller as the more zeros appearing in the simulation. Therefore, the value of precision with varying volatility is less instructive in this study. Based on this, we categorize the outcomes in two parts according to two sizes of precision, the large precision with the value from 3 to 4 and the small precision with the value from 1 to 1.5. No matter in the simulation with large or small precision, the result indicates that the percentage of the NPI interval including the benchmark value gets lower, and the NPI results are less accurate along with the increasing volatility. If the results are compared horizontally, it is not difficult to conclude that the NPI prediction presents a better result with a larger precision than that with a smaller precision. Thus, inputting a larger precision is a safer choice. From the finance perspective, it illustrates that a conservative investor who uses NPI method can do the prediction with a larger precision to behave safely, but at the same time he may miss a lot of trading chances in the market.  In the simulation with a large precision, when the volatility is lower than 4.5\%, the NPI method can offer a good prediction with the percentage greater than 80\%, and accuracy less than 2. To get a better result with a percentage greater than 90\% and accuracy less than 1.5, the volatility needs to be restricted within 3\%.  In the simulation with small precision, the NPI's result is good when the volatility is also lower than 4.5\%, and  the corresponding accuracy is less than 2.08 worse than the one with large precision. But if a better result is required, the volatility should be lower than 2.5\% in order to make the percentage greater than 90\%, then the accuracy under these circumstances is less than 1.61. Overall, we can draw the conclusion that the NPI method performance is better with an option based on an underlying asset at a lower volatility less than 3\% daily. An Asian option is normally used in commodity and foreign exchange markets where the underlying asset is less volatile than the equity in the stock market. This allows the NPI method to provide a relatively good result for the Asian option pricing in these markets. To support the statement, an empirical example of the Asian option in the crude oil market is investigated.

\subsection{The empirical study of the energy market}
\begin{figure}
\centering
\includegraphics[scale=0.8]{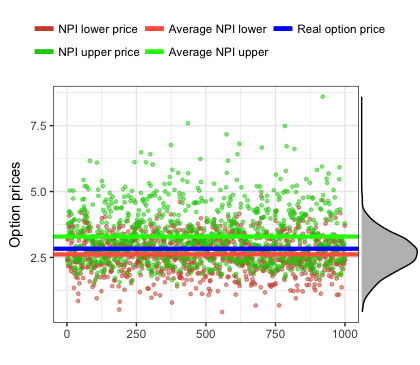}
\caption{NPI predictions with a ten year historical data }\label{fg:WTILarge}
\end{figure}  

The crude oil commodity market is considered in this example. The set of data is the New York Mercantile Exchange (NYMEX) daily closed price of  the WTI crude oil normally used as a benchmark in the oil pricing. The Asian option price is the Chicago Mercantile Exchange (CME group) the WTI average price call option started on 23/10/2019 and expired on 29/11/2019 with the strike price 54(\$). By the end of the trading time on 23/10/2019, the settlement price of this call option is 2.83(\$), which is a reference price provided by the CME group. The NPI method, an imprecise statistical framework based on the historical data, controls the precision of the prediction by managing the historical data size. By large historical data, the degree of prior information dispersion  is more significant than a small historical data leading to a less precise interval result. In the following example, we forecast the average price option based on ten years of historical data, from 23/10/2009 to 23/10/2019, and the plotted result is shown in Figure \ref{fg:WTILarge}. We first calculate the daily volatility based on the historical data, which equals to 2.1\%. According to our volatility study by simulation, the NPI prediction is supposed to provide some valuable results under this volatility. After 10000 trails, we get the expected NPI maximum buying price is equal to 2.6152 and the expected NPI minimum selling price is equal to 3.29124, which are shown as the two horizontal lines in orange and green in Figure \ref{fg:WTILarge}. The dots in Figure \ref{fg:WTILarge} the NPI results in each trail. To make the graph clear and well recognized, only 1000 trails results are plotted in the figure. Figure \ref{fg:WTILarge} indicates that the NPI method provides a relatively good result that includes the real price in its interval. 

 \begin{figure}
\centering
\includegraphics[scale=0.8]{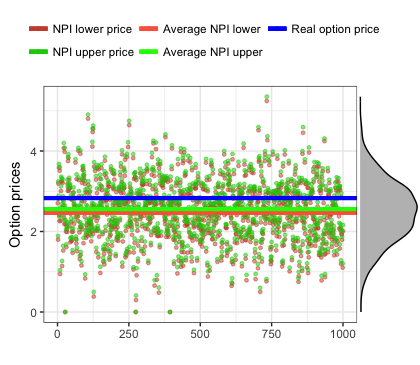}
\caption{NPI predictions with a one year historical data}\label{fg:WTISmall}
\end{figure}
 
 Next, we improve the precision of the NPI prediction by limiting the historical data size to one year, from 23/10/2018 to 23/10/2019. From Figure \ref{fg:WTISmall}, the trial outcomes are more concentrated leading to a more precise result with a smaller interval from the NPI method. The maximum buying price is 2.447181, and the minimum selling price is 2.538937. The daily volatility during this historical period is 2.4\%. Although the NPI result is more precise, the interval deviates distinctly from the real market price 2.83 comparing to the result from a large historical data. This means we scarify the accuracy in order to gain a more precise result. But this is not a good deal since the investor referring to the NPI result would lose money when he trades WTI in the market. 
 
 \begin{figure}
\centering
\includegraphics[scale=0.8]{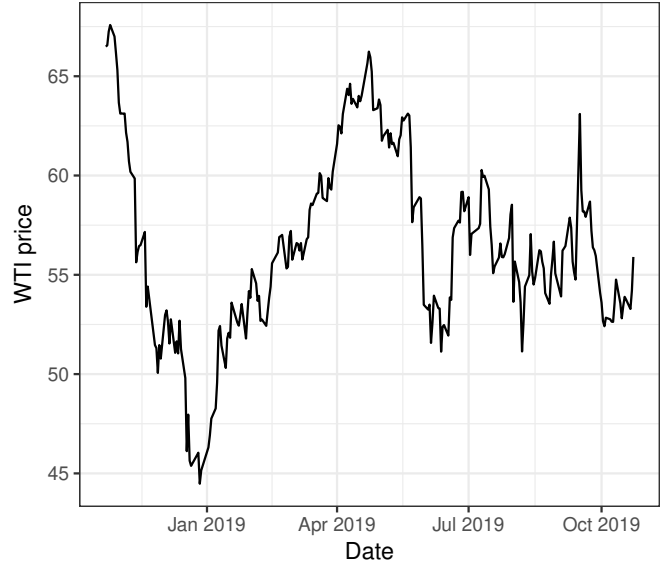}
\caption{WTI crude oil price from 23/10/2018 to 23/10/2019 }\label{fg:WTIPrice}
\end{figure}

\begin{figure}
\centering
\includegraphics[scale=0.8]{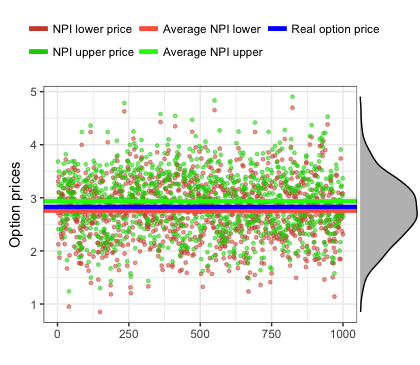}
\caption{NPI predictions after adjustment}\label{fg:WTISmallAd}
\end{figure}

The examples above are a rigid investigation of the NPI performance by controlling the size of historical data. To study the historical data more clearly, we display the WTI crude oil price in this recent one year in Figure \ref{fg:WTIPrice}. It is obvious that there is a deep drop that started in October 2018 ended in December 2018, which is the worst performance in nearly three years. The price is down to 44.48 on 27/12/2018 the lowest closing price since January 2016. There exist multiple reasons causing this drop, global oversupply keeping the investors away, investors with less confidence of economic recovery in the next year and the longest US government shutdown on 22 December 2018. Through the comprehensive consideration, the data from 23/04/2019 to 23/10/2019 has a better reference value to do the prediction. But it is an arbitrary decision to cut off the data of early half year crudely. What we do here is to adjust the sampling procedure making it focus more on the latter half years' data than the earlier one. To achieve this, we  use the maximum and minimum one year historical prices as the boundary values, but the main sampling data is the historical data from 23/04/2019 to 23/10/2019. By doing this, the pricing procedure not only considers the probabilities of the unexpected event but also places emphasis on the historical information in a relatively stable market environment. The adjusted result is plotted in Figure \ref{fg:WTISmallAd}. It is obvious that after adjustment, the accuracy of the NPI result gets better, the maximum buying price at 2.757054 and the minimum selling price 2.936124. This interval covers the real market price which is a better investment indicator than the NPI result without adjustment. In addition, the precision of the NPI result is nearly as same as that without adjustment. This example manifests that the NPI method performs better combining with the assessment of historical data.

\begin{table}
\centering
\begin{tabular}{c||cccc}
  \hline
  \\[0.3ex]
Trading Date & $\overline{P}_{\text{WTI}}$ & $\underline{P}_{\text{WTI}}$ & $\overline{P}_{\text{Brent}}$ & $\underline{P}_{\text{Brent}}$\\
  \hline
2019-11-22& 0.15 & 0.13 & 0.04 & 0.03 \\ 
2019-11-23 & 0.88 & 0.87 & 0.27 & 0.25 \\ 
  2019-11-24& 0.80 & 0.79 & 0.17 & 0.15 \\ 
  2019-11-25 & 0.70 & 0.69 & 0.04 & 0.03 \\ 
  2019-11-26 & 0.70 & 0.69 & 0.04 & 0.02 \\ 
  2019-11-27 & 0.71 & 0.69 & 0.03 & 0.02 \\ 
  2019-11-28 & 0.71 & 0.70 & 0.03 & 0.01 \\ 
  2019-11-29 & 0.72 & 0.71 & 0.04 & 0.03 \\ 
 2019-11-30& 0.97 & 0.96 & 0.41 & 0.40 \\ 
   \hline
\end{tabular}
\caption{NPI probabilities for WTI vs Brent with $K=0.95*S_0$}\label{ta:WTIB}
\end{table}

\begin{figure}
\centering
\includegraphics[scale=0.8]{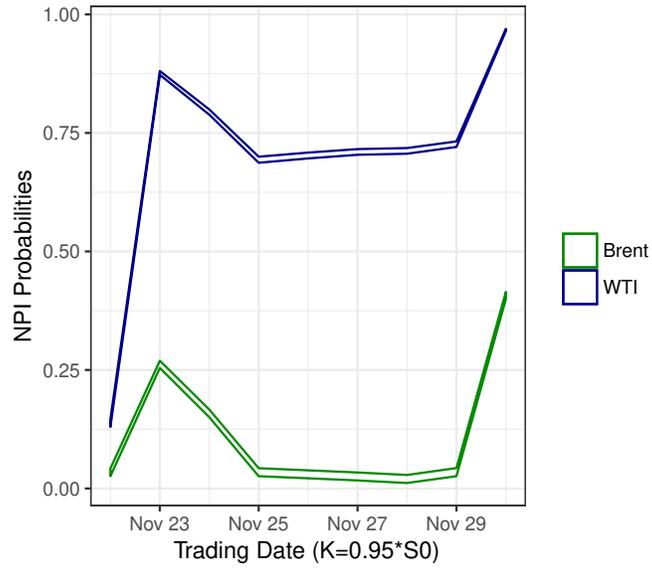}
\caption{NPI probabilities for WTI and Brent}\label{fg:WTIB}
\end{figure}

The NPI method as discussed in Section \ref{s:2.2} can be used as the market director for an investor. As acknowledged, in the crude oil market, WTI from the American and Brent  from the North Sea are two benchmark prices of the crude oil market that are both sweet and normally track one another. Their prices trend and pattern are similar to each other making the investor hard to compare these two values directly from the market price. According to the NPI method illustrated in Section \ref{s:2.2}, an investor can get an indicator of the trading action according to the investor's risk aversion. In the following example, we assume the investor wants to buy an Asian call option on either WTI or Brent that the average underlying asset price during the option life period is not less than 95\% of its spot price meaning the strike price $K$ is equal to 95\% of the spot price $S_0$. The call option's trading day is from 2019-11-22 to 2019-11-30, and the expire day is 2019-11-30, so the option period is from 9 days to 0 days. Then the upper and lower probabilities of both WTI and Brent average prices  greater than strike price are calculated getting the results displayed in Table \ref{ta:WTIB}. From Table \ref{ta:WTIB}, it is obvious that the lower probabilities of the WTI price are greater than the upper probability of the Brent price.  WTI definitely has a higher possibility to earn a positive payoff in the call option market. For the investor either as a speculator or a hedger, it is optimal to invest in WTI. The result also plotted in Figure \ref{fg:WTIB} showing that the probability pattern of the WTI price is similar to that of the Brent price but with greater values. Also, from the figure, we can tell the best time to get in the market, which is 2019-11-23 in this example, since the NPI probabilities of Nov 23rd are the greatest value among these dates except the one of Nov 30th.  The WTI and Brent oil price  returns from 2019-11-22 to 2019-11-30 are also calculated to assess our prediction. The WTI return equals to 4.102\% higher than the Brent return, 1.935\%, which confirms that the trading strategy based on NPI is profitable.

 \begin{figure}
\centering
\includegraphics[scale=0.8]{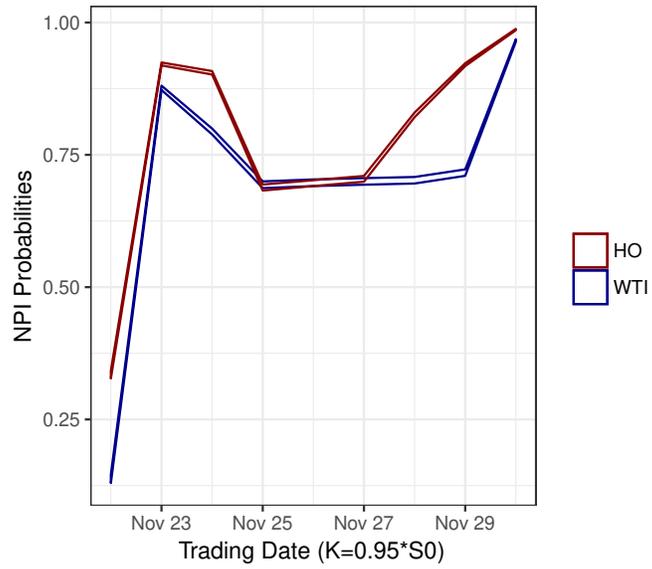}
\caption{NPI probabilities for WTI and the heating oil}\label{fg:WTIH}
\end{figure}

\begin{table}
\centering
\begin{tabular}{c||cccc}
  \hline
  \\[0.3ex]
Trading Date & $\overline{P}_{\text{WTI}}$ & $\underline{P}_{\text{WTI}}$ & $\overline{P}_{\text{HO}}$ & $\underline{P}_{\text{HO}}$\\
  \hline
2019-11-22 & 0.15 & 0.13 & 0.34 & 0.33 \\ 
  2019-11-23 & 0.88 & 0.87 & 0.92 & 0.92 \\ 
  2019-11-24 & 0.80 & 0.79 & 0.91 & 0.90 \\ 
  2019-11-25 & 0.70 & 0.69 & 0.69 & 0.68 \\ 
  2019-11-26 & 0.70 & 0.69 & 0.70 & 0.69 \\ 
  2019-11-27 & 0.71 & 0.69 & 0.71 & 0.70 \\ 
  2019-11-28 & 0.71 & 0.70 & 0.83 & 0.82 \\ 
  2019-11-29& 0.72 & 0.71 & 0.92 & 0.92 \\ 
  2019-11-30& 0.97 & 0.96 & 0.99 & 0.99 \\ 
   \hline
\end{tabular}
\caption{NPI probabilities for WTI vs Heating Oil (HO) with $K=0.95*S_0$}\label{ta:WTIH}
\end{table}
To end this study, we calculate the upper and the lower probabilities of the WTI price comparing to another oil product, the heating oil. The heating oil price is related to the WTI price, because it is  a low viscosity, liquid petroleum product  made from the WTI crude oil. So the price of the heating oil in the United States is depending on the supply of the WTI crude oil. Here we also pick the WTI and heating oil price data from the CME group during 2019-11-22 to 2019-11-30. The event of interest is the average price $S_\mu$ ended by Nov 30th is greater than the  95\% of the spot price $S_0$. From the perspective of the Asian option, we are interested in the probability of a call option with $K=0.95*S_0$ end up with a positive payoff. We plot the NPI upper and lower probabilities in Figure \ref{fg:WTIH}. The decision of the option selection is harder to make in this comparison  group, because there are intersections and overlapping of the NPI probabilities between these two products. Unlike the result of WTI versus Brent that WTI always dominates, in this example, there are overlapping and intersections in Figure \ref{fg:WTIH}. The different underlying asset is picked according to the trading date. To specify the underlying asset selection based on the trading data, the exact value of NPI upper and lower probabilities are listed in Table \ref{ta:WTIH}.  Between 2019-11-22 to 2019-11-24, the lower probability of $S_\mu >K$ for the heating oil  dominates the upper probability of $S_\mu>K$ for  WTI. During this period, a speculator or a hedger is better to invest in the heating oil. On Nov 25th, the upper and lower probabilities of WTI are greater than the corresponding value of the heating oil. So on this day, a speculator is supposed to choose the call option based on the WTI oil price, but a hedger would wait. On Nov 26th,  the NPI probability intervals of WTI and the heating oil are overlapped with each other, so there is no indication which underlying asset is better. The next day's upper probabilities of these two oil price are still the same value, while the lower probability of  WI is less than the lower probability of the heating oil. Thus, on Nov 27th, a speculator is better to get in the game of the call option based on the heating oil, and a hedger still waits for the sign of a more determined trading indicator. This indicator appears on Nov 28th and lasts until Nov 30th, the lower probability of the heating advantages over the upper probability of the WTI, leading to the trade for both a speculator or a hedger in the Asian call option contingent on the heating oil.

\section{Concluding Remarks}
This paper presents a novel approach to evaluate the Asian option with the arithmetic price from the imprecise probability aspect through the NPI method, which forecasts the option price on the basis of the historical data with few assumptions. This approach provides an interval of prices as the result, which not only contains the uncertainty from the probability perspective but also the uncertainty from limited prior information. This property makes  it more advanced than the traditional method for the Asian option, especially the one in the energy market because of the less liquidity of the Asian option in the energy market. The NPI method also gives a risk measure by comparing two energy products inspiring the investor with a trading criterion. We study the performance of the NPI method first by the simulation using the GBM prediction as the benchmark. Three factors, precision, coverage percentage and accuracy are defined and investigated to help us assess the performance. It turns out the NPI forecasting has more reference value for the less volatile product. Then we predict the WTI crude oil price based on the NPI method comparing to the real market price. With a long period of historical data, the NPI forecasting interval contains the real market price, but the precision of the result is not entirely satisfactory. To get a more precise interval result, narrowing the size of the historical data is going to scarify the accuracy of the result. After the investigation, we found that using the extreme value of the historical data to control the precision and considering the historical event of adjusting the sampling period of the historical data can offer a better outcome. We also illustrate the risk measure, the NPI trading criterion, by two examples, the trade of  WTI and Brent and the trade of WTI and the heating oil. An investor is guided according to their risk aversions by using this criterion.

In order to get a better result from the NPI method, there are several aspects we need to consider. The time period of the predictive data should be considered discreetly since we assume the exchangeability of all data including the historical data and the future data in Section 2. If the prediction period is too long, it challenges the reasonableness of the exchangeability assumption, because some significant fluctuation may happen in the market. To inference these fluctuations, a large historical data is needed to infer the situation, which as we discussed in the last paragraph, this will reduce the level of accuracy of the result. The extreme value of the historical data, $r_{(0)}$ and $r_{n+1}$, is also an important consideration that will affect the NPI prediction as we illustrated in the example. These two values play a very important role in balancing the precision of the result and the inference ability of the historical data for the significant fluctuations. 

To avoid the effect from the unexpected historical event or seasonal effect, the sampling historical data $s_{(1)},...,s_{n}$ has been picked in the time period with a more stable market.  However, dealing with these effects may be important, the adaption of the NPI method for the data with the seasonal effect is a meaningful topic for future study. In this paper, we have explained that the NPI method is more suitable to predict the price of a less volatile product. How to solve the prediction problem of a market with high volatility from the imprecise probability perspective is also appealing for future study.

\bibliographystyle{cas-model2-names}

\bibliography{cas-refs}

\section*{Data  avaliability statement}
The raw data that support the study in this paper is obtained from the CME websites and the S\& P Capital IQ database.

\end{document}